# Using Off-Resonance Laser Modulation for Beam Energy Spread Cooling in Generation of Short-Wavelength Radiation

Haixiao Deng

*Shanghai Institute of Applied Physics, Chinese Academy of Sciences, Shanghai, 201800, P. R. China*

Various seeding configurations have being proposed for frequency up-conversion of the electron beam density distribution, in which the energy spread, may however hinder the harmonic generation efficiency. In this Letter, a method for cooling the electron beam energy spread by off-resonance seed laser modulation is described, using a transversely dispersed beam and a modulator undulator with proper transverse gradient. With this novel mechanism, it is shown that the frequency up-conversion efficiency can be significantly enhanced. We present theoretical analysis and numerical simulations for seeded soft x-ray free electron laser and storage ring based coherent harmonic generation in extreme ultraviolet spectral region.

Email: denghaixiao@sinap.ac.cn

Availability of high brightness and short-wavelength radiations, especially x-ray pulses, is of great interest. X-ray pulses enable the simultaneous probe of both the ultra-small and the ultra-fast worlds, and continuously revolutionize the understanding of the matters. Therefore, synchrotron radiation (SR) light sources and free-electron lasers (FELs) based on advanced particle accelerators are being developed worldwide to satisfy the dramatically growing demands in the material and biological sciences [1]. The fundamental process of SR and FEL sources usually involves a relativistic electron beam passing through a transverse periodic magnetic field, e. g. the undulator, and generating electromagnetic radiation ranging from the infrared to hard x-ray regions, depending on the electron beam energy and the undulator period and strength. More recently, the successful user operation of the first FEL facilities [2-5] in soft and hard x-ray regimes announced the birth of x-ray laser. Currently, the light source community is on the stage to more sophisticated and determined schemes, e.g., in pursuit of fully coherence [6-18], fast polarization switch [19-22] and compact x-ray configurations [23-26].

Coherence describes all properties of the correlation between physical quantities of a single wave, or between several waves or wave packets. It is widely used in any field that involves waves, such as acoustics, electrical engineering, neuroscience, and quantum mechanics. Spatial coherence describes the correlation between waves at different position, which is naturally ensured in SR and FEL sources. Temporal coherence describes the predictable relationship between waves observed at different times, which is usually obtained by purifying the noisy spectra with crystal monocharomator in SR sources. As the leading lasing mode of the hard x-ray FEL, self-amplified spontaneous emission (SASE) [27] starts from the initial shot noise of the electron beam and results in radiation with poor temporal coherence. Recently it is demonstrated in the liner coherent light source [3] that, the temporal coherence of SASE can be significantly improved by the configuration of self-seeding [6] and iSASE [9].

Alternatively, in order to generate fully coherent radiation, various seeded FEL schemes [10-18] were proposed and intensively studied around the world. One of the most initial seeded FEL configurations is high gain harmonic generation (HGHG) [10], which has been perfectly demonstrated in the visible and ultraviolet region [11-12]. Unfortunately, the standard-HGHG suffers an essential drawback that a single stage allows only a limited frequency multiplication factor. Thus, multi-stage HGHG approach [13-14] was proposed for short-wavelength production from an ultraviolet seed wavelength, which leads a rather complication in the overall design. Meanwhile, novel concepts are under development, e.g., echo-enabled harmonic generation (EEHG) [15] which has been experimentally demonstration at the 3$^{rd}$, 4$^{th}$ and 7$^{th}$ harmonics [16-18], in principle could efficiently work at several tens of harmonic of in a single stage.

The goal of this Letter is to point out a new physical mechanism that has important advantages over the classical approach to the frequency up-conversion. As we will show below, the energy spread will be cooled effectively by the transversely dispersed electron beam passing through a modulator undulator with proper transverse gradient, thus holds promising results in generating of short-wavelength radiation. Using theoretical analysis and numerical simulations, it is demonstrated that with energy modulation amplitude of 7 times of the initial energy spread, the bunching factor of the 30$^{th}$ harmonic of the seed laser could be up to 15%. In order to clearly illustrate the setup for the new mechanism, we will quickly review the traditional way to modulate the beam current using a seed laser and a modulator undulator tuned to the laser frequency, i.e., the standard-HGHG.

The standard-HGHG setup is shown in Fig. 1(a). Generally, the pulse length of the electron beam is much larger than the seed laser wavelength, and the beam current variation over the distance of several laser wavelengths can be neglected. Thus, we locally assume a longitudinally uniform electron beam. And Gaussian distribution beam energy with the energy spread $\sigma$ is applied. The external laser and the electron beam interact with each other in the modulator undulator. Then after passing through the dispersive chicane, the electron beam's energy modulation is converted into density modulation, which ensures the temporally coherent at high harmonic of the seed laser and can be characterized as the bunching factor,

$$b_n = e^{-\frac{n^2 D^2 \sigma^2}{2}} J_n(nD\Delta\gamma) \tag{1}$$

Where $\Delta\gamma$ is the maximum energy modulation at the end of the modulator undulator, $n$ is the interested harmonic order, $D$ is the dispersive strength of the chicane and $J_n$ is the Bessel function of order $n$. It follows from Eq. (1) that the bunching factor $b_n$ decays exponentially with $n$ increase because of the presence of the beam energy spread [10, 28-29].

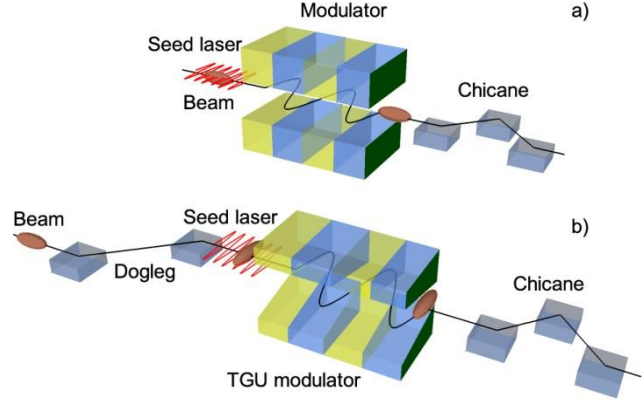

FIG. 1 (color online). A standard-HGHG system (a) consists of a modulator undulator and a dispersive section. The proposed cooled-HGHG scheme (b) includes a dogleg, a modulator undulator with transverse gradient and a chicane.

To overcome the low efficiency of the standard-HGHG to modulate the beam, we propose to use an off-resonance seed laser modulation for cooling down the electron beam energy spread. The setup is depicted in Fig. 1(b), which we call cooled-HGHG. After transversely dispersing the electron beam with average energy $\gamma_0 mc^2$ by the dogleg dispersion $\eta$, it is injected to a modulator undulator with transverse gradient of $\alpha$ and central dimensionless parameter of $K_0$, then one can give a linear $x$ dependence of the undulator parameter

$$K(x) = K_0\left(1 + \alpha\eta \frac{\gamma - \gamma_0}{\gamma_0}\right). \tag{2}$$

For the interested wavelength of the seed laser, the resonant beam energy should be written as

$$\gamma_r(x) = \gamma_0 + \alpha\eta \frac{K_0^2}{K_0^2 + 2}(\gamma - \gamma_0). \tag{3}$$

Now we consider the longitudinal dynamics of two electrons $(\gamma_0, \theta_0)$ and $(\gamma, \theta_0)$. At the exit of the modulator undulator, the two electrons becomes $(\gamma_0', \theta_0)$ and $(\gamma', \theta_0)$. And for the small $\theta_0$, one has

$$\begin{cases} \gamma_0' = \gamma_0 + \Delta\gamma\sin\theta_0 = \gamma_0 + \Delta\gamma\theta_0 \\ \gamma' = \gamma + \Delta\gamma\sin(\theta_0 + \Delta\varphi) = \gamma + \Delta\gamma(\theta_0 + \Delta\varphi) \end{cases}, \tag{4}$$

where $\Delta\varphi$ is the phase advance of the off-resonance electron with respect to the on-resonance electron, which can be given by the phase equation of electron in modulator undulator,

$$\Delta\varphi = 4\pi N \frac{(\gamma - \gamma_r)}{\gamma_0}. \tag{5}$$

Where $N$ represents the period number of the modulator, combing Eq. (4) and Eq. (5), we can easily derive that

$$\frac{\gamma' - \gamma_0'}{\gamma - \gamma_0} = 1 - \frac{4\pi N \Delta\gamma}{\gamma_0}\left(\frac{\alpha\eta K_0^2}{K_0^2 + 2} - 1\right). \tag{6}$$

Eq. (6) illustrates a universal scaling of the off-resonant laser modulation as a tool of electron beam energy spread cooling. It is clear that, in standard-HGHG setup, the beam energy spread is amplified by a factor of $4\pi N\Delta\gamma/\gamma$ which is usually a relatively small number. When we increase the product of $\alpha\eta$ and make the second term of the right hand in Eq. (6) to be zero, the beam energy spread is maintained because it is fully compensated by the undulator transverse gradient as in ref. [24]. If one further increases the product of $\alpha\eta$ properly, there exists a pattern of minimum electron beam energy spread, thus the frequency up-conversion efficiency should be dramatically enhanced according to the Eq. (1). For a given energy modulation, the bunching factor is nearly determined by the maximum of the Bessel function, which can be optimized by the longitudinal dispersion of Chicane in Fig. 1(b). One should note that since the beam energy modulation is induced by the seed laser in the modulator undulator, the transverse achromatic of the whole system is not necessary.

In order to clearly illustrate the essence of the cooled-HGHG, GENESIS1.3 [30] simulation using tiny emittance and tiny beam size, i.e., one-dimensional (1D) simulation was carried out on the basis of the parameters of the Shanghai soft x-ray FEL test facility (SXFEL) [31]. The baseline design of SXFEL is a two-stage HGHG from 265nm to 8.8nm. The parameters used here are: slice energy spread of 84keV, beam energy of 0.84GeV, modulator period length of 80mm and period number of 12. Since the frequency multiplication factor in the first stage of SXFEL is 6, we introduce a maximum energy modulation of 600keV by the 265nm seed laser in the modulator, i.e., 7 times of the initial beam energy spread. Then a two-dimensional scan of the $30^{th}$ harmonic bunching factor, as a function of $\eta$ and $\alpha$ was carried out. The result in Fig. 2 demonstrates a stable optimal zone with maximum bunching factor over 20%. It is found that the optimal relation is approximately $\alpha\eta \approx 24$ from the simulation, while Eq. (6) indicates an optimal formula of $\alpha\eta = 12$. One should note that in the cooled-HGHG, a factor of 1/2 should be introduced in Eq. (5), because the energy spread approximately decreases monotonously in the modulator undulator, thus the phase advance $\Delta\varphi$ obtained by integration over the modulator undulator contributes a factor of 1/2 [32]. Considering many rough assumptions in the theoretical analysis, it is reasonably consistent with the simulation results.

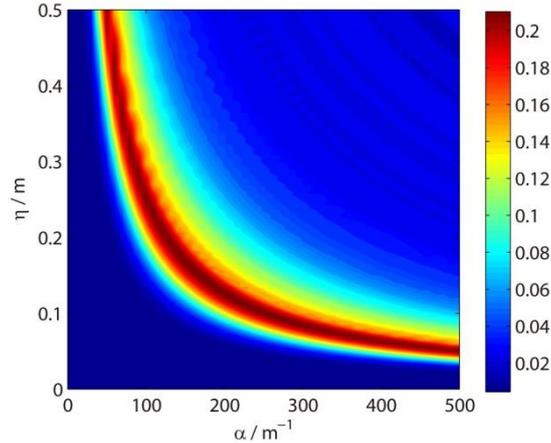

FIG. 2: (color online) Optimization of the transverse gradient $\alpha$ of the modulator and the transverse dispersion $\eta$ of the dogleg by 1D simulation, in order to find the optimal bunching factor of the $30^{th}$ harmonic for the cooled-HGHG.

With the optimized parameters above, Fig. 3 shows the evolution of the phase space of the beam as it travels through the standard-HGHG and the proposed cooled-HGHG. These pictures demonstrate a simple physical mechanism behind the off-resonance laser modulation effect. Because of the electron beam energy spread cooling indicated in Eq. (6), after the off-resonance laser modulation, it leads to a merger phenomenon around the center of the longitudinal beam phase space, and a split effect at the two flanks. Considering that most of the electrons are concentrated around the center, the density modulation will be significantly enhanced in the cooled-HGHG.

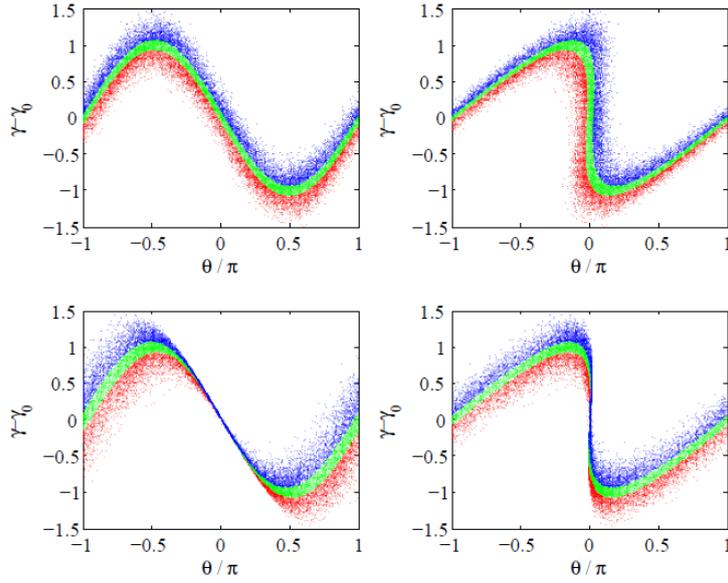

FIG. 3: (color online) The phase space after the modulator undulator (top left) and the dispersive chicane (top right) in the standard-HGHG, and the phase space after the modulator undulator (bottom left) and the dispersive chicane (bottom right) in the cooled-HGHG. The blue, green and red represent the electrons with high, medium and low energy at the beginning of the modulator undulator, respectively.

The above results are corroborated by the data shown in Fig. 4, where the maximum achievable bunching at different harmonic is plotted. Fig. 4 also shows the results from the standard-HGHG for comparison purpose. These results were calculated by optimizing the longitudinal dispersion for each harmonic while using the phase space in Fig. 3. While the bunching factor exponentially decreases as the harmonic number increases in the standard-HGHG, it can be well maintained in the cooled-HGHG scheme, e.g., over 15% even at the 50th harmonic.

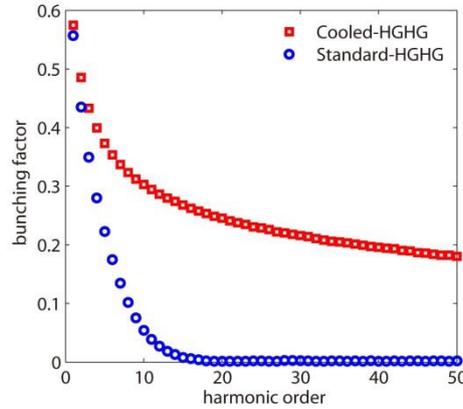

FIG. 4 (color online). Bunching factor vs harmonic number at the exit of the dispersive chicane in the standard-HGHG and the cooled-HGHG configuration.

The theoretical estimates and numerical findings obtained using the simple 1D model have been checked by means of the 3D simulation, which properly takes into account the seed laser diffraction and the transverse beam dynamics. The remarkable performance of the cooled-HGHG was fully confirmed by the 3D simulation in Fig. 5. The bunching factor of the $30^{th}$ harmonic of the seed laser exceeds 15%. Further study demonstrates that this strong bunching drives an initial quadratic growth in the followed undulator tuned at 8.8nm, and the peak power saturates above 0.5GW after a 12m long undulator. It is worth stressing that, since the cooled-HGHG benefits from the linearly transverse dispersion of the beam in the modulator undulator, a large ratio of the beam size contributed by the dogleg dispersion to the $\beta$ function is preferred. Thus in the 3D simulation, 0.8mm laser radius, 0.1μmrad normalized emittance and 10μm electron beam size in horizontal plane were assumed.

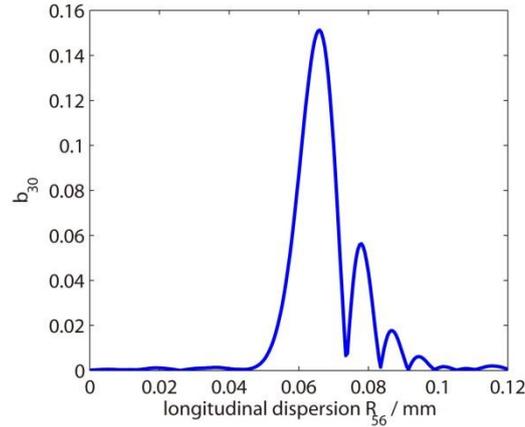

FIG. 5: (color online) The bunching factor of the 30$^{th}$ harmonic as a function of the dispersive strength of chicane for the cooled-HGHG, from the 3D simulation result.

Now we consider the coherent harmonic generation setup on synchrotron radiation facilities, also known as optical klystron, in which the large energy spread is the main obstacle on the way to short-wavelength. We take the UVSOR-II storage ring parameters as an example, i.e., 0.6GeV beam energy, 0.6MeV energy spread, total geometry emittance of 17.5nmrad, coupling of 3%. Considering the small beam emittance in the vertical plane, the electron beam was proposed to be vertically dispersed, and 800nm laser together with 20 periods' modulator undulator induce a maximum energy modulation of 1.2MeV. Then with an optimal relation of $\alpha\eta \approx 6.5$ from the 3D simulation, the bunching factor of the 6$^{th}$ harmonic is enhanced to 23.0% in the cooled HGHG from 1.8% in the traditional setup, and thus the intensity of the 133nm extreme ultraviolet radiation will be enhanced by two orders in the followed undulator. It should be pointed out that, one could introduce a smaller energy modulation to get sufficient density modulation at an interested wavelength by using the cooled-HGHG, which will effectively avoid the radiation pulse stretch by the microwave instability, and thus offer a better time resolution for users.

In summary, we proposed a new mechanism to remarkably cool the electron beam energy spread by off-resonance laser modulation. It is demonstrated with theoretical analysis and numerical simulations that the proposed technique holds great prospects in frequency up-conversion based on high brightness LINACs and storage rings. There are several practical physical effects that were not included in these simple considerations. They include the method of generating the required dispersion, coherent and incoherent synchrotron radiation effects in dispersive elements, the transverse gradient imperfection of the modulator undulator. These and other effects should be taken into account before carrying out a proof-of-principle experiment. Moreover, it is worth stressing that the proposed mechanism can be easily extended to other applications, e.g., substituting the first stage of an echo-enabled harmonic generation by the proposed one will further enhance the density modulation performance.

The author would like to thank Meng Zhang, Dazhang Huang and Haohu Li for beam dynamics issue, Bo Liu, Dong Wang and Yuantao Ding for enthusiastic discussions on FEL physics and TGU simulations, Chao Feng for help in the draw design and critical remarks. This work was partially supported by the Major State Basic Research Development Program of China (2011CB808300) and the National Natural Science Foundation of China (11175240 and 11205234).